\documentstyle[hip-artc]{article}
\volnumber{9}  \edyear{1999}  \frompage{000} \topage{000}                
\recrevdate{1 January 1999}                                              
\def\rightmark{Sudden freeze-out  vs  continuous  emission: duality...}
\def\leftmark{S.V. Akkelin et al.}

\title{Sudden freeze-out  vs  continuous  emission: duality  in hydro-kinetic
approach  to A+A  collisions}
\authors{
{\twerm S.V. Akkelin$^1$, M.S. Borysova$^{2}$ and Yu.M.
Sinyukov$^1$ }\\[2.812mm] {\normalsize \hspace*{-8pt}$^1$
Bogolyubov Institute for Theoretical Physics, \\  03143 Kiev,
Metrologichna 14b, Ukraine \\[0.2ex]
\hspace*{-8pt}$^2$ Taras Shevchenko National University, \\ 01033
Kiev, Volodymirs'ka 64, Ukraine }}

\abstract{The problem  of spectra  formation in hydrodynamic
approach  to  A+A collisions is  discussed.  It is analyzed in
terms of the two different objects: distribution and emission
functions. We show that though the  process  of particle
liberation, described by the emission function, is, usually,
continuous in time, the observable spectra can  be also expressed
by  means of the Landau/Cooper-Frye prescription. We argue that
such an approximate duality results from some symmetry properties
that systems in A+A collisions reach to the end of  hydrodynamic
evolution and reduction of  the collision rate at post
hydrodynamic stage.} 

\begin{document}

\maketitle
\section{Introduction}

The two main features of high energy nucleon and nuclear
collisions, namely, strong interactions and multiparticle
production, suggest an idea to Heisenberg, Watagin and Fermi
\cite{Heisenberg} that the systems created in those collisions
could be considered as thermal. The Landau hydrodynamic model for
multi-hadron production \cite{Landau} appears as a method that is
alternative to S-matrix one, where the asymptotic states at $t=\pm
\infty $ are not considered but instead the detail space-time
evolution of thermal hadronic matter is described basing on the
local energy-momentum conservation laws. Instead of the initially
well defined asymptotic state of colliding particles, this
approach uses initial conditions of hydrodynamic expansion that,
in fact, strongly depend on pre-thermal stage of hadronic or
nuclear collisions. At the RHIC energies the pre-thermal stage is
formed in collisions of the two groups of specifically distributed
partons associated with colliding nuclei that results, apparently,
in very dense state of partonic matter: Color Glass Condensate
(see, e.g., \cite{CGC}). Current phenomenological and pQCD
estimates give the (proper) times $\tau _{0}$ of thermalization of
that matter between 0.6 and 3 fm/c. Different phenomenological
assumptions are used to fix the details of thermalization; well
known examples are Landau initial condition: compressed static
disk at $t=\tau _{0}$ \cite{Landau} and Bjorken one:
quasi-inertial flow at proper time $\tau =(t^{2}-z^{2})^{1/2}=\tau
_{0}$ \cite{Bjorken}.

The another important aspect of the hydrodynamic approach is the
final state of colliding system. If one uses hydrodynamic
equations till infinitely large times, the result will be the
infinitesimally small final densities that gives no chance to
restore microscopic (or particle) picture from continuous medium
approach. According to the Landau criterion of freeze-out, the
hydro-evolution stops and all particles become free when fluid
elements reach the temperature that is equal to the mass of the
lightest hadron (pion): $T_{f.o.}\equiv 1/\beta _{f.o.}=m_{\pi }.$
At that temperature the mean free path in pion gas becomes to be
equal to transverse radius of hydrodynamic tube in
$p+p(\overline{p})$ collisions and system decays.  This
qualitative criterion is in mysterious qualitative agreement with
hydrodynamic
descriptions of very wide class of high energy collisions: from $p+p(%
\overline{p})$ to heavy ion A+A (see, e.g., \cite{xu}) where the
correspondent fitting freeze-out temperature turns out to be
120$\div $150 MeV. This universality seems like a puzzle since it
does not take into account the real dynamics and system sizes.
Moreover, the results of many studies of A+A collisions based on
cascade (transport) models contradict to an idea of sudden
freeze-out at some fixed temperature. The particles escape from
the system during the whole period of its evolution and do not
demonstrate the local equilibration at the late stages. Though the
pure hadronic cascade models as well as hybrid ''hydro + cascade''
models \cite{hybrid} fail to describe properly experimental data,
especially the HBT radii in A+A collisions, the problem of spectra
formation in the hydrodynamic approach in itself is very serious
and only stress the puzzle of the successful application of the
hydrodynamic models to multiparticle production processes.

\section{Distribution and emission functions and observables}

To clarify the problems we will follow to the basic ideas of Ref.
\cite{Sinyukov} and express the observables through both
distribution and emission functions of expanding systems. We start
from the Boltzmann equations (BE) as more general than the
hydrodynamic ones. The BE for the distribution function $f(x,p)$
in the case of no external forces has the form:
\begin{equation}
\frac{p^{\mu }}{p^{0}}\frac{\partial f(x,p)}{\partial x^{\mu }}%
=F^{gain}(x,p)-F^{loss}(x,p).  \label{Boltzmann}
\end{equation}
The term $F^{gain}$ and $F^{loss}$ are associated with the number
of particles which respectively came to the point $(x,p)$ and
leave this point because of collisions. The term
$F^{loss}(x,p)=R(x,p)f(x,p)$ can easily be expressed in terms of
the rate of collisions of the particle with momentum $p $,
$R(x,p)=\,<\!\sigma v_{rel}\!>n(x)$. The term $F^{gain}$ has more
complicated integral structure and depends on the differential
cross-section. Let us split the
distribution function at each space-time point into two parts: $%
f(x,p)=f_{int}(x,p)+f_{esc}(x,p)$, $x=(t,\mathbf{x})$. The first one, $%
f_{int}(x,p)$, describes the fraction of the system which will continue to
interact after the time $t$. The second one, $f_{esc}(x,p)$, describes the
particles that will never interact after the time $t$. According to the
probability definition
\begin{equation}
f_{esc}(x,p)=P(x,p)f(x,p),  \label{prob-def}
\end{equation}
where escape probability $P(x,p),$ or probability for any \textit{%
given} particle at $x$ with momentum $p$ not to interact any more,
propagating freely, can be expressed explicitly in terms of the rate of
collisions along the world line of the free particle with momentum $p$
through the opacity integral
\begin{equation}
P(x,p)=\exp \left( -\int\limits_{t}^{\infty }dt^{\prime
}R(x^{\prime },p)\right)   \label{prob-calc}
\end{equation}
Since $f_{esc}(x,p)$ is formed from the particles suffering last
collisions at space-time point $x$ it is associated with term $P
F^{gain}$ and form the emission function $S$ \cite{Sinyukov}
\begin{equation}  \label{eq-f+}
p^\mu \frac \partial {\partial x^\mu }f_{esc}(x,p)=p^0P%
(x,p)F^{gain}(x,p)\,\equiv S(x,p).
\end{equation}

For initially finite system with a short-range interaction among particles,
the system becomes free, in fact, at large enough times $t_{out}$, so $%
P(x,p)\rightarrow 1$ and $f_{esc}(x,p)\rightarrow f(x,p)$ in this
limit. Therefore, to describe the inclusive spectra of particles
\begin{equation}
p^{0}\frac{dN}{d\mathbf{p}}=\langle a_{p}^{+}a_{p}\rangle \,,\
p_{1}^{0}p_{2}^{0}\frac{dN}{d\mathbf{p\mathtt{_{1}}}d\mathbf{p}\mathtt{_{2}}}%
=\langle a_{p_{1}}^{+}a_{p_{2}}^{+}a_{p_{1}}a_{p_{2}}\rangle ,
\label{spectra-def}
\end{equation}
the asymptotic equality $f_{esc}(x,p)=f(x,p)$ can be used,
replacing the total distribution function $f$ in all irreducible
averages in (\ref{spectra-def}),
\begin{equation}
\langle a_{p_{1}}^{+}a_{p_{2}}\rangle =\int_{\sigma _{out}}d\sigma
_{\mu }p^{\mu }\exp {(iqx)}f(x,p),  \label{average-wigner}
\end{equation}
by $f_{esc}.$ Here, $p=(p_{1}+p_{2})/2\,$, $q=p_{1}-p_{2}$ and the
hypersurface $\sigma _{out}$ just generalizes $t_{out}\,$.
Applying the Gauss theorem and recalling that $\partial _{\mu
}[p^{\mu }exp{(iqx)}]=0$ for particles on mass shell, one obtains,
using respectively general equations (\ref{eq-f+}) and
(\ref{Boltzmann}) and supposing their analytical continuation off
mass shell,
\begin{eqnarray}
\langle a_{p_{1}}^{+}a_{p_{2}}\rangle  &=&p^{\mu }\int_{\sigma
_{0}}d\sigma _{\mu }f_{esc}(x_{0},p)e^{{iqx}}+\int_{\sigma
_{0}}^{\sigma _{out}}\!d^{4}x\,S(x,p)e^{{iqx}},\hspace*{-0.3cm}
\label{sp-e} \\ \langle a_{p_{1}}^{+}a_{p_{2}}\rangle  &=&p^{\mu
}\int_{\sigma _{0}}d\sigma _{\mu
}f(x_{0},p)e^{{iqx}}+p^{0}\!\!\int_{\sigma _{0}}^{\sigma
_{out}}\hspace*{-0.3cm}d^{4}x(F^{gain}(x,p)\!-\!F^{loss}(x,p))e^{{iqx}},
\label{sp-f}
\end{eqnarray}
where $S(x,p)$ is defined through the product $PF^{gain}$  by  Eq.
(\ref{eq-f+}) ,  $f_{esc}(x_{0},p)$  corresponds to the portion of
the particles, which is already free at initial time $t_{0}$, or,
more generally, at the initial hypersurface $\sigma _{0}$, and
$f(x_{0},p)$ is the distribution function at $\sigma _{0}$.

Thus, the use of escaping function as the asymptotic interpolation
to the solution of BE is equivalent to taking, as the source
function for the spectra and correlations, the 4-volume emission
function $S=p^{0}PF^{gain}$ together with direct emission
$f_{esc}(x_{0},p)$ from an initial 3D hypersurface $\sigma _{0}$.
The Landau criterion of freeze-out of locally equilibrium (l.eq.)
hydrodynamic momentum spectra and correspondent Cooper-Frye
prescription, defined by Eq. (\ref{average-wigner}) with substitutions $%
\sigma _{out}\!\rightarrow \sigma _{f.o.}$ and $f\!\rightarrow
f_{l.eq.}$, treats particle spectra as results of rapid conversion
of a l.eq. hadron system into a gas of free particles at some
hypersurface $\sigma _{f.o.}$. Formally, it corresponds to taking
the cross-section tending to infinity at $t<t_{\sigma _{f.o.}}$
(to keep system in l.eq. state) and zero beyond $t_{\sigma
_{f.o.}}$. Then $P(t,{\bf x},p)=\theta (t-t_{\sigma
_{f.o.}}\!(\mathbf{x}))$ (and so $f_{esc}=0$ at $t<t_{\sigma
_{f.o.}}$), and $S=p^{0}PF^{gain}=p^{0}\delta (t-t_{\sigma
_{f.o.}}\!({\bf x}))f_{l.eq.}$ in Eq. (\ref{sp-e}). The
proportionality between the $S$ and $f_{l.eq.}$, like $S(x,p)=\rho
(\tau )f_{l.eq.}(x,p)$ is used in many papers
 devoting to a description of the data in a hydrodynamically
motivated way (see review \cite{Heinz}). \textit{One should
understand, however, that in the realistic case of no sudden
freeze-out the emission function } $S(x,p)$
\textit{loses completely its proportionality to the distribution function} $%
f(x,p)$, \textit{it is just two different objects! }This was shown
by direct calculations in Ref. \cite{Sinyukov} and also is clear
from Eqs. (\ref{prob-def}), (\ref{eq-f+}) since the escape
probability $P(x,p)$ in finite systems is very sensitive to an
asymmetry of the positions $\mathbf{x} $ as for the effective
''boundary'' and, thereby, the $S(x,p)$ becomes anisotropic in
momentum $\mathbf{p}$ in the rest frame of a fluid element unlike
to the hydrodynamic distribution function $f_{l.eq.}(x,p)$.

\section{The duality of the hydro and kinetic approaches}

As it was advocated in the recent papers \cite{Shuryak} the
perfect hydrodynamics is a good approximation for the earlier
stage of the matter evolution because of the big cross-section of
the interaction among colour and white quasi-hadronic states in
QGP. As for the matter evolution at the post hadronization stage
in A+A collisions it was found the approximate equivalence between
chemically frozen hydrodynamics of hadron gas and its evolution
within the cascade approach in the temperature region above 0.12
GeV \cite{Teaney}. Below this region the formation of spectra is
continuous in time with fairly long  ''tails'' of the emission.
That process is characterized by the emission function $S(t,{\bf
x,p}) $ that is far from thermal. The similar situation was
analyzed in Ref. \cite{Sinyukov} based on the exact analytic
solution  of the BE for expanding fireball. Unlike to very
complicated structure of the emission function $S(t,{\bf x,p}) $
that was found in \cite{Sinyukov}, the spectra, interferometry
radii, averaged phase-space densities have simple and clear
analytical forms and correspond to the Cooper-Frye prescription
for the  thermal distribution function $f_{l.eq.}(x,p)$ in the
fireball \textit{before} it starts to decay. The explanation is
based on the duality of Eqs. (\ref{sp-e}) and (\ref{sp-f}) that
express the spectra in terms of either the emission or
distribution functions. While the emission function, that is
proportional to $F^{gain}$, can have a significant non-zero value
in wide space-time region, the integral of the difference
$F^{gain}-F^{loss}$ over this region could be zero, as in the
example discussed in Ref. \cite{Sinyukov}, or small. The latter is
typical when the system  expands in spherically symmetric way. The
analysis of different hydrodynamic solutions  demonstrates that
the velocity field of expanding systems tends typically to a
spherically symmetric one at the late stage of  evolution, at
least, in the central region where  the low $p_{T}$ spectrum
forms. As for the high $p_{T}$ spectrum formation, one can expect
that correspondent particles are radiated mainly from the
periphery of the system at the earlier times, because of large
hydrodynamic velocities and fast transition to free streaming
there: it was argued in Refs. \cite{Sinyukov} and
\cite{freeze-out}. Thereby, in a rough approximation, one can
apply the generalized Cooper-Frye prescriptions (\ref{sp-f}) putting there $%
\sigma _{0}=\sigma _{f.o.}$, taking into account the possible
$p$-dependence of hypersurface $\sigma _{f.o.}(p)$ at high
$p_{T}\succeq 0.8-1$ GeV and neglecting the integral over the 4D
region situated beyond of the $\sigma _{f.o.}(p)$. That is some
kind of duality in the description of the spectra and the
interferometry data basing either on (thermal) distribution functions $%
f_{l.eq.}(x,p)$ which characterize the system just before decay
begins or on the emission function $S(t,{\bf x,p})$ that describes
the process of continuous particle liberation during the decay  of
system.

In fact, any kind of hydrodynamic or Boltzmann kinetic approaches
loses its applicability when the particle mean free paths become
compatible with lengths of homogeneity in the system.
The universality of the Landau freeze-out temperature, $%
T\simeq m_{\pi }$, an independence on the concrete dynamics and
correspondent homogeneity lengths  is due to the pions are
dominated in hadron system and their mean free paths, $1/\sigma
n(T(x))$, start to increase exponentially  $\sim \exp (\beta
m_{\pi })$ with $\beta =1/T(x)$ when the temperature falls down
below $m_{\pi }$ as it follows from the analytic representation of
the thermal density $n(T)$. Thus the temperature $T\simeq m_{pi}$
is just lower boundary of the region of applicability of
hydrodynamics in wide class of nucleon and nuclear collisions. It
does not mean that the hadrons stop to interact then at post
hydrodynamic stage but the momentum spectra do not change
significantly especially if the above discussed conditions (small
value  of  $ F^{gain}$, symmetry of expansion at the late stage,
etc.) are satisfied and so the integral of $F^{gain}-F^{loss}$ in
Eq. (\ref{sp-f})  is small at that stage.

\section{Conclusions and outlook}

We show  that universal Landau freeze-out temperature corresponds
 to lower boundary  of  the applicability of
hydrodynamics that  is  similar  in  different  collisions. The
duality in  spectra description between the (generalized) sudden
freeze-out prescription, that utilizes the distribution functions
$f$, and the detail picture of the particle liberation process,
based on the emission function $S$, is argued. The replace of the
complicated emission process by the simple Landau \cite{Landau}
criterion of sudden freeze-out at $T\approx 0.12\div 0.15$ GeV is,
of course, a rather rough approximation. Nevertheless, as it was
discussed in Ref. \cite{AkkSin}, the momentum-energy conservation
laws, peculiarities of almost isoentropic and chemically frozen
evolution as well as some symmetry features of  the late stage of
hydro evolution minimize the correspondent uncertainties. Note,
that the duality does not  mean and even excludes the
parametrization  of emission  function in the form $%
S\sim $ $f_{l.eq.}$  excepting  for  the  case  of  real  sudden
freeze-out  which  is, probably,  very  non-realistic.

The approximation discussed is, unfortunately, not  well
controlled in practical utilization and we hope to develop an
approximate method for spectra calculations within the
hydro-kinetic approach \cite{Sinyukov} based on the escape
probabilities and generalized relaxation time approximation. The
method will combine the advantages of hydrodynamic approximation
and microscopic (kinetic) approach. The former allows one to
incorporate the complicated evolution of the system at the
possible phase transitions encoded in corresponding equation of
state; the latter makes possible to evaluate the observable
spectra taking into account the non-equilibrated character of
their formation. As a result we hope coherently explain all
totality of hadronic observables of various experiments where
existing dynamic models fail to accommodate the majority of
experimental data.
\section*{Acknowledgments}

The work was supported by NATO Collaborative Linkage Grant No.
PST.CLG.980086.

\vfill\eject
\end{document}